\documentclass{kluwer}    
\usepackage{graphicx}
\newdisplay{guess}{Conjecture}

\begin{document}       
\begin{article}

\begin{opening}         
\title{CDM, Feedback and the Hubble Sequence}
\author{Jesper \surname{Sommer-Larsen}, Martin \surname{G\"otz} and Laura
\surname{Portinari}}
\runningauthor{Sommer-Larsen, G\"otz \& Portinari}
\runningtitle{CDM, Feedback and the Hubble Sequence}
\institute{Theoretical Astrophysics Center, Copenhagen, Denmark}
\begin{abstract}
We have performed TreeSPH simulations of galaxy formation in a standard
$\Lambda$CDM cosmology, including effects of star formation, energetic stellar
feedback processes and a meta-galactic UV field, and obtain a mix of disk, 
lenticular and elliptical galaxies.
The disk galaxies
are deficient in angular momentum by only about a factor of two compared to
observed disk galaxies. 
The stellar disks have approximately exponential surface density profiles, and
those of the bulges range from exponential to $r^{1/4}$ , as observed.
The bulge-to-disk ratios of the disk galaxies are consistent
with observations and likewise are their integrated $B-V$ colours, which have
been calculated using stellar population synthesis techniques.  
Furthermore, we can match the observed $I$-band Tully-Fisher (TF)
relation, provided that the mass-to-light ratio of disk galaxies is
$(M/L_I) \sim$ 0.8.
The ellipticals and lenticulars have approximately $r^{1/4}$ stellar surface
density profiles, are dominated by non-disklike kinematics and flattened due to
non-isotropic stellar velocity distributions, again consistent with 
observations.

\end{abstract}
\keywords{Cosmology, galaxy formation, numerical methods}
\end{opening}           
\vspace{-10mm}
\section{Introduction}
\vspace{-5mm}
The hierarchical Cold Dark Matter (CDM) structure formation scenario has
proven remarkably successful on large (cosmological) scales. On galactic 
scales it has encountered a number of problems, most notably the angular
momentum problem, the over-cooling problem, the missing satellites problem
and the central cusps problem. Inclusion of the effects of energetic stellar 
feedback processes in galaxy formation simulations may help to cure a number
of these problems as indicated by, e.g., the ``toy'' models of Sommer-Larsen
et al. (1999, SLGV). Recently Sommer-Larsen \& Dolgov (2001, SLD) showed that 
by going from the
CDM structure formation scenario to warm dark matter (WDM) scenarios one can
alleviate and possibly even completely overcome the angular momentum 
problem, and complementary work of Colin et al. (2000) shows this to be the 
case also for other of the above problems. Fine-tuning of the warm dark matter 
particle mass to about 1 keV is
required, however. In contrast the salient feature about ``conventional'' CDM 
is that as long as the dark matter particles are much heavier than one keV, 
the actual particle mass does not matter for structure formation.

We have recently completed a series of considerably more elaborate CDM
galaxy formation simulations with very encouraging results: We find that    
a mix of disk, lenticular and elliptical galaxies 
can be obtained in fully cosmological ($\Lambda$CDM), gravity/hydro 
simulations invoking star-formation, energetic stellar feedback processes and 
a meta-galactic UV field. These results, together with results on disk gas
infall histories and stellar age distributions, hot halo gas properties,
global star formation histories etc. are presented in detail in 
Sommer-Larsen et al. (2002).   
\vspace{-5mm}

\section{The simulations}
\vspace{-5mm}
Star-formation efficiencies are assumed to be related to the thermal history
of the gas effectively making the efficiencies large ($\sim$1) at early times
($z >$ 4--5) and much smaller ($\sim$0.01) at later times. Early star-formation
is assumed to be either a) self-propagating or b) only ``local'', the former 
resulting in the strongest bursts and feedback events. 
The early, fast star formation is assumed to be triggered when the gas density
of an SPH particle exceeds a certain critical value, chosen to be 
$n_{\rm{H,fast}}=0.3$ cm$^{-3}$. 
The conversion of such an SPH particle into a star particle may or may not
trigger a burst of self-propagating star formation (SPSF) in the cold, dense 
gas
surrounding it: in scenario a) not only the SPH particle which gets above
the critical density threshold, but also its neighbouring cold and dense
SPH particles with densities above $n_{\rm{H,fast,low}} (< n_{\rm{H,fast}})$
are triggered for conversion into star particles on their individual, 
dynamical timescales. Such SPSF is observed in some 
star-burst galaxies (e.g., in expanding super-shells --- see Mori et al.
1997). In scenario b) only the initial SPH particle above the critical density
threshold is triggered for star formation on the dynamical time scale.

We selected 12 dark matter halos from a cosmological $\Lambda$CDM N-body 
simulation
for the galaxy formation simulations. The masses of these halos spanned
more than a factor of 10 and their characteristic velocities $V_{200}$
range from 100 to 250 km/s. After resampling the galaxy formation simulations
consisted of 30000-150000 SPH+DM particles. We started out by running all
12 galaxy simulations using the SPSF prescription
with a lower density threshold of
$n_{\rm{H,fast,low}}$=0.1 cm$^{-3}$.
Seven of the resulting galaxies at $z$=0 had distinctly disk galaxy like
morphologies and kinematics, the remaining 5 lenticular (S0) or elliptical
like morphologies and kinematics. 4 additional series of simulations were
subsequently run for the 7 disk galaxies: Three using again early 
SPSF with $n_{\rm{H,fast,lower}}$=0.05, 0.2 and 0.25 
cm$^{-3}$ and one series with fast, early, but non-SPSF. The four choices of 
$n_{\rm{H,fast,lower}}$ results in conversion of
2--5\% of the gas in the simulations into stars in the early bursts. In the
models without SPSF about 1\% of the gas is turned into stars in the early 
bursts.
\vspace{-10mm}
\section{Results}
\vspace{-5mm}
The disk galaxies have 
the bulk of their stars on 
approximately circular orbits in a disk, most of the rest of the stars 
in an inner, bulge-like component and finally a small fraction in a round and 
dynamically insignificant stellar halo surrounding the galaxies. The disk
galaxies formed in our simulations are hence qualitatively quite similar to
the Milky Way and other disk galaxies. Of the remaining 5 galaxies, two
have a minor fraction of the stars on nearly circular, disk orbits; we classify
these as lenticulars (S0s), and the remaining three have no stars at
all on disk like orbits; we classify these as ellipticals.

The disk galaxies have approximately exponential stellar disk surface density
profiles and exponential to $r^{1/4}$ bulge profiles, all in good agreement
with observations. The lenticular and elliptical galaxies are characterized by
approximately $r^{1/4}$ stellar profiles.

The bulges of the disk galaxies are generally confined to being within 
$r_B \sim 1-1.5$ kpc from the centers of the galaxies. Bulge-to-disk ratios
were determined by extrapolating the nearly exponential disk profiles outside
of $r_B$ to the center of the galaxies. Using these decompositions (which
make no assumptions about the bulge surface density profiles) the specific 
angular momenta of the stellar {\it disks} were estimated taking explicitly
into account also the region with overlap between disk and bulge. 
Characteristic circular speeds $V_c$ for the disk galaxies were calculated
using the approach of SLD, but as an addition taking into account also the 
dynamical effect of the bulges.

\vspace{-7mm}
\begin{figure}[h]
\leavevmode
\includegraphics[width=10.5truecm]{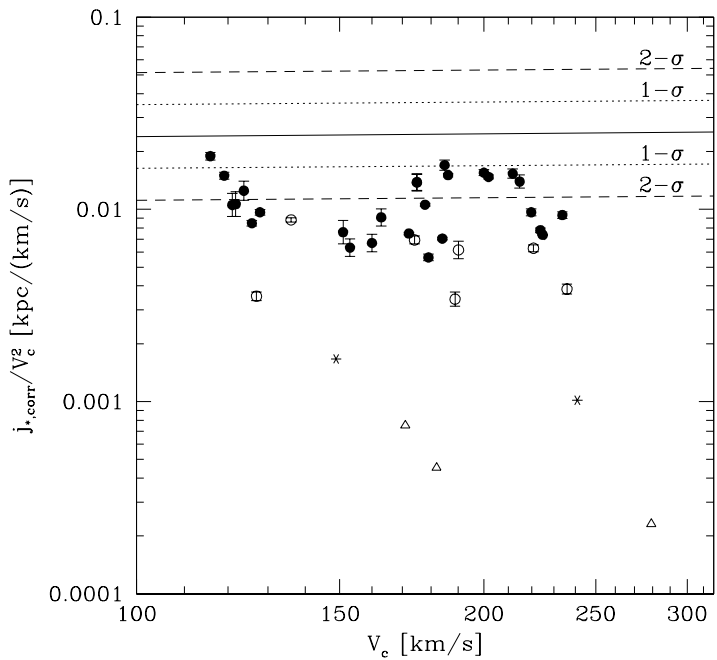}
\vspace{-8mm}
\caption{Normalized specific angular momenta for all galaxies.
Filled circles: disk galaxies formed in simulations with SPSF, open circles: 
disks formed in simulations without SPSF, 
star symbols: S0's and triangles: E's.}
\end{figure}
\newpage
\vspace{-10mm}
\begin{figure}[h]
\leavevmode
\includegraphics[width=10.5truecm]{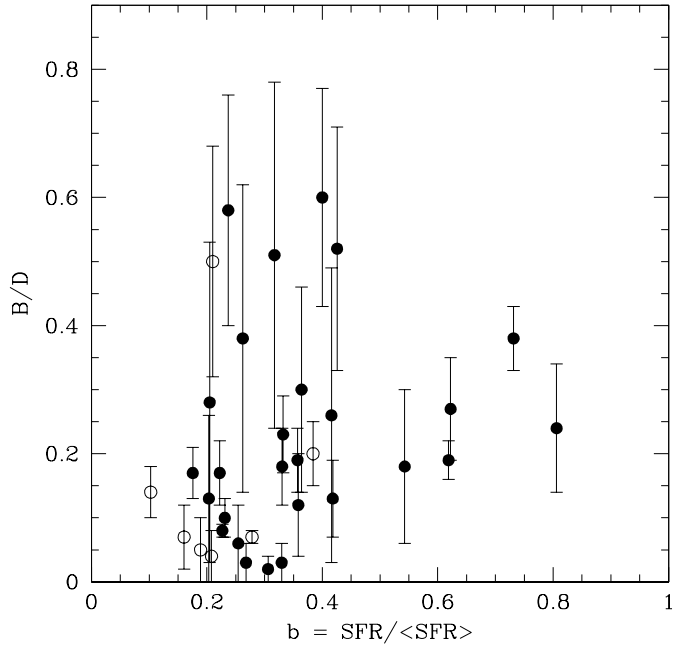}
\vspace{-8mm}
\caption{Stellar bulge-to-disk ratios for the 35 disk galaxies --- symbols
as in Fig. 1.}
\end{figure}
\vspace{-2mm}
In Figure 1 we show the ``normalized'' specific angular momenta 
$\tilde{j}_{*} = j_{*}/V_c^2$ of the final disks formed in all
35 disk galaxy simulations as a function of $V_c$. 
As argued by SLGV one expects $\tilde{j}_{disk}$ to 
be almost independent of $V_c$ on both theoretical
and observational grounds. Also shown in the figure is the median 
``observed'' value of $\tilde{j}_{disk}$, calculated as in SLGV and SLD
for a Hubble parameter $h$=0.65, together with the observational 
1-$\sigma$ and 2-$\sigma$ limits. As can be seen from the figure, the
specific angular momenta of the stellar disks from the SPSF simulations
lie only about a factor of two
below the observed median (the specific angular momenta of the disks from the 
SPSF simulations have been spin-parameter corrected --- see SLD).
This is about an order of magnitude better than
what is obtained in similar CDM simulations without energetically effective, 
stellar feedback processes, as discussed by many authors, and almost as good
as was obtained by SLD for WDM. The simulations without SPSF and the 
associated, strong feedback events do not do as well.
Also shown in the figure are the normalized specific angular momenta of the
two lenticular and three elliptical galaxies. These
are about an order of magnitude smaller than those of the disk galaxies, 
broadly consistent with observations.

Figure 2 shows the bulge-to-disk ratios $B/D$ of the 35 disk galaxies versus
the birthrate parameter $b$, which is the ratio of the current to the
average past star formation rate. $b$ is a disk galaxy type indicator, with
$b \sim 0.1$ for Sa's increasing to $b \sim 1$ for Sc's (Kennicutt, Tamblyn \&
Congdon 1994) 
The trend (or rather, lack of trend) seen in Figure 2 is
broadly consistent with observationally 
determined (and 2-D decomposed) $I$ and $K$-band $B/D$s, which 
trace the {\it mass} bulge-to-disk ratios fairly well - see Sommer-Larsen et 
al. (2002).

\vspace{-17mm}
\begin{figure}[h]
\leavevmode
\includegraphics[width=10.5truecm]{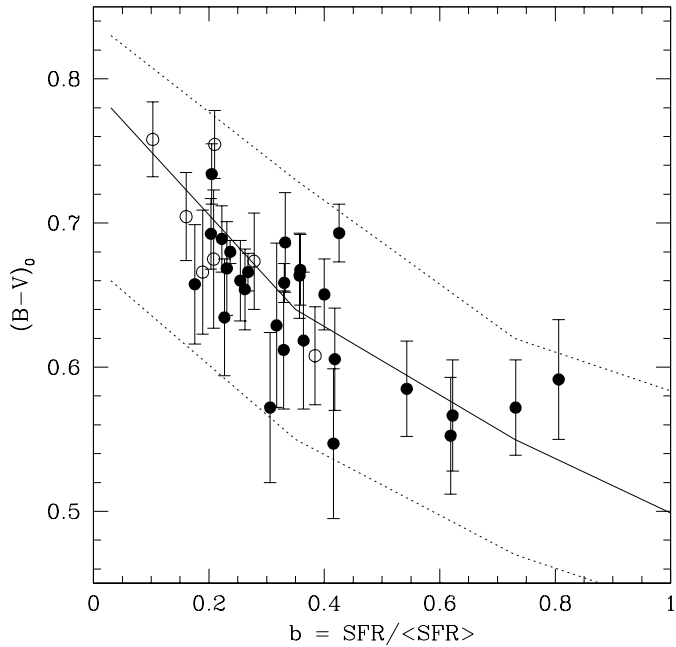}
\vspace{-3mm}
\caption{Integrated $B-V$ colours for the 35 disk galaxies, calculated using
stellar population synthesis techniques --- symbols as in Fig.1.
Also shown (solid line) are the mean 
observational values for disk galaxies from Roberts \& Haynes (1994) and 
the observational 1-$\sigma$ limits (dotted line).} 
\end{figure}

Integrated $B-V$ colours were obtained for the galaxies by stellar population
synthesis techniques. In Fig. 3 we show these for the 35 disk galaxies vs.
the $b$ parameter. Also shown are observations --- the general agreement is
excellent.
Finally, in Fig. 4 we show $M_{*}(V_c)$ of the final disk galaxies formed in 
35 runs together with the $I$-band Tully-Fisher relation (TF) for $h$=0.65, 
converted
to mass assuming mass-to-light ratios $(M/L_I)$ = 0.5, 1.0 and 2.0. The
slope of the ``theoretical'' TF matches that of the observed very well
for a constant mass-to-light ratio, which is required to be $(M/L_I) 
\sim$ 0.8, similar to the findings of SLD for their WDM simulations. 
Such a low value is consistent with recent, dynamically estimated 
mass-to-light ratios for disk galaxies,
the mass-to-light ratio of the Milky Way (Fig. 4) and can be obtained from 
stellar
population synthesis models provided an IMF somewhat less ``bottom-heavy'' 
than the Salpeter law is used (Portinari, Sommer-Larsen \& Tantalo 2002,
Sommer-Larsen et al. 2002). 

\vspace{-10mm}
\begin{figure}[h]
\leavevmode
\includegraphics[width=10.5truecm]{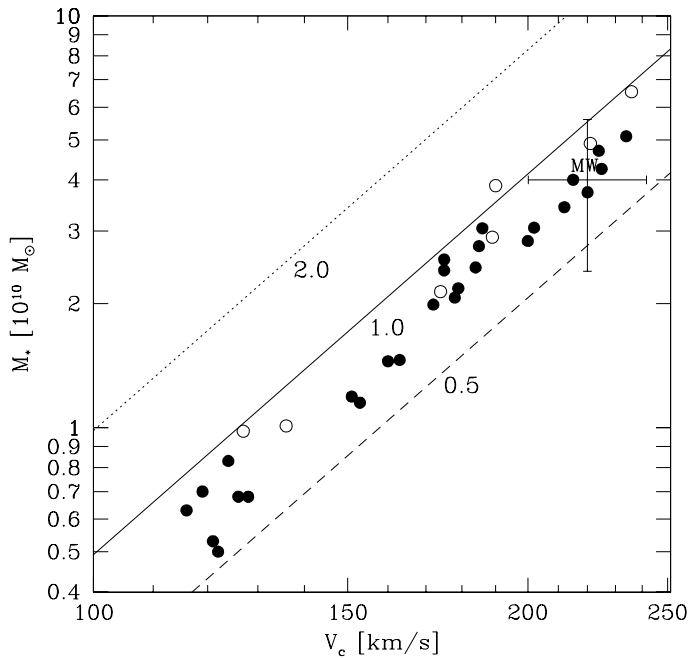}
\vspace{-3mm}
\caption{The stellar {\it mass} ``Tully-Fisher'' relation for the 35 disk
galaxies --- symbols as in Fig.1. Error bars marked MW: The stellar mass of the
Milky Way. Lines: observed TF relation for $(M/L_I)$ = 0.5, 1.0 
and 2.0 - see Sommer-Larsen et al. (2002) for details.}
\end{figure}
\vspace{-6mm} 
\section*{Acknowledgement}
\vspace{-4mm}
We thank the organizers for an in all aspects truly outstanding 
conference.

\end{article}

\vspace{-6mm}
\begin{thebibliography}{}

\vspace{-3mm}
\bibitem{}
Colin, P., Avila-Reese, V.\ and Valenzuela, O., 2000, ApJ, 542, 622
\bibitem{}
Giovanelli, R., et al., 1997, ApJ, 477, L1
\bibitem{}
Kennicutt, R., Tamblyn, P.\ and Congdon, C., 1994, ApJ, 435, 22
\bibitem{}
Mori, M., Yoshii, Y., Tsujimoto, T.\ Nomoto, K., 1997, ApJ, 478, L21
\bibitem{}
Portinari, L., Sommer-Larsen, J. and Tantalo, R., 2002, in preparation
\bibitem{}
Roberts, M.\ and Haynes, M., 1994, ARA\&A, 32, 115
\bibitem{}
Sommer-Larsen, J.\ and Dolgov, A., 2001, ApJ, 551, 608 (SLD)
\bibitem{}
Sommer-Larsen, J., Gelato, S.\ and Vedel, H., 1999, ApJ, 519, 501 (SLGV)
\bibitem{}
Sommer-Larsen, J., G\"otz, M.\ and Portinari, L., 2002, ApJ, submitted
(astro-ph/0204366)
\end{thebibliography}
\end{document}